\documentclass[10pt, aps, prd, superscriptaddress, nofootinbib, showpacs, twocolumn]{revtex4-2}
\usepackage[english]{babel}
\usepackage{amsmath, amsfonts, amsthm, amssymb, mathrsfs}
\usepackage[all]{xy}
\usepackage{bm}
\usepackage{stmaryrd}
\usepackage{graphicx}
\usepackage[colorlinks]{hyperref}
\usepackage{color}
\usepackage{subcaption}
\usepackage{amscd}

\usepackage{pgfplots, grffile, tikz}
\pgfplotsset{compat=newest}
\usetikzlibrary{plotmarks, arrows.meta}
\usepgfplotslibrary{patchplots}

\usepackage{dsfont}

\numberwithin{equation}{section}

\usepackage{stmaryrd}
\usepackage{mathtools}

\newcommand{\lb}{\left[}
\newcommand{\rb}{\right]}

\renewcommand{\a}{\alpha}

\newcommand{\g}{\gamma}
\renewcommand{\d}{\delta}

\newcommand{\s}{\sigma}

\newcommand{\lbar}{\lower0.4ex\hbox{$\mathchar'26$}\mkern-9.5mu \lambda}

\renewcommand{\geq}{\geqslant}

\begin{document}

\title{Schwinger--DeWitt expansion for the heat kernel of nonminimal operators in causal theories}

\author{A. O. Barvinsky}
\email{barvin@td.lpi.ru}
\affiliation{Theory Department, Lebedev Physics Institute, Leninsky Prospect 53, Moscow 119991, Russia}

\author{A. E. Kalugin}
\email{kalugin.ae@phystech.edu}
\affiliation{Theory Department, Lebedev Physics Institute, Leninsky Prospect 53, Moscow 119991, Russia}

\author{W. Wachowski}
\email{vladvakh@gmail.com}
\affiliation{Theory Department, Lebedev Physics Institute, Leninsky Prospect 53, Moscow 119991, Russia}

\begin{abstract}
We suggest a systematic calculational scheme for heat kernels of covariant nonminimal operators in causal theories whose characteristic surfaces are null with respect to a generic metric. The calculational formalism is based on a pseudodifferential operator calculus which allows one to build a linear operator map from the heat kernel of the minimal operator to the nonminimal one. This map is realized as a local expansion in powers of spacetime curvature, dimensional background fields, and their covariant derivatives with the coefficients---the functions of the Synge world function and its derivatives. Finiteness of these functions, determined by multiple proper time integrals, is achieved by a special subtraction procedure which is an important part of the calculational scheme. We illustrate this technique on the examples of the vector Proca model and the vector field operator with a nondegenerate principal symbol. We also discuss smoothness properties of heat kernels of nonminimal operators in connection with the nondegenerate nature of their operator symbols.
\end{abstract}

\maketitle

\section{Introduction}
Here we suggest a systematic approach to the solution of a long standing problem---the generalization of Schwinger--DeWitt expansion \cite{DeWitt1965} to heat kernels of nonminimal operators. As is well known, this asymptotic expansion at $\tau\to 0$ for the two-point heat kernel of the matrix-valued differential operator $F=\hat F(\nabla)$,
\begin{multline} \label{DW}
\hat K_F(\tau|x,x') =  e^{-\tau\hat F(\nabla_x)} \delta(x,x')\\
=\frac{\Delta^{\frac12}(x,x')}{(4\pi\tau)^{\frac{d}2}}\, g^{\frac12}(x')\,
e^{\textstyle{-\frac{\sigma(x,x')}{2\tau}}}\! \sum\limits_{m=0}^\infty \tau^{m}\,\hat a_m(F\,|\,x,x'),
\end{multline}
underlies ultraviolet (UV) renormalization (and more generally, local effective field theory expansion) of models in curved spacetime with generic background fields.\footnote{We work in $d$-dimensional spacetime with the Euclidean signature metric $g_{ab}(x)$ and fiber bundle of fields $\varphi=\varphi^A(x)$ having generic indices, $A=1,2,...M$, acted upon by $M\times M$ matrices or matrix-valued operators, denoted by hats, $\hat X\equiv X^A_B$, $\hat 1\equiv\delta^A_B$. $\sigma(x,x')$ is the Synge world function and $\Delta(x,x')$ is its dedensitized Pauli-VanVleck determinant, $g(x)={\rm det}\,g_{ab}(x)$.} The virtue of this expansion is that it incorporates a systematic way to calculate coincidence limits $\hat a_m(F\,|\,x,x)$ which are the celebrated HaMiDeW or Gilkey--Seeley \cite{Gilkey1975,Gilkey1995,Barvinsky1985,Scholarpedia,Vassil03} coefficients---local polynomials in spacetime curvature and background fields associated with the operator $F$.

However, this expansion is true only for covariant {\em minimal} second-order operators $\hat F(\nabla)$ of the form
\begin{equation} \label{minimal}
\hat F(\nabla) = -\Box\,\hat 1 + \hat P.
\end{equation}
Here $\Box=g^{ab}\nabla_a\nabla_b$ is a covariant d'Alembertian defined in a curved spacetime, $\hat P=\hat P(x)$ is a potential term. Covariant derivatives $\nabla_a$ act on fiber bundle fields and preserve $g_{ab}$, $\nabla_c g_{ab}=0$. This formulation includes the case of extra first-order derivative term in $\hat F(\nabla)$, which can be generated by the extension of the fiber bundle connection, $\nabla_a\to\nabla_a+\hat\varGamma_a$.

The expansion (\ref{DW}) can be adjusted to higher order minimal operators \cite{Gusynin1990,Wach3,BKW2024} whose leading in derivatives term is given by the power of the covariant d'Alembertian, $\hat H(\nabla)=(-\Box)^N+...$, $N>1$, but for the nonminimal operators in physical applications there exist only indirect calculational methods bypassing the construction of the heat kernel, like the method of universal functional traces of \cite{JackOsborn1984,Barvinsky1985}.

In contrast to minimal operators the {\em nonminimal} ones,
\begin{equation}\label{nonminimal_operator}
    \hat H(\nabla)=\hat D^{a_1...a_{2N}}\nabla_{a_1}...\nabla_{a_{2N}}+O\big(\,\nabla^{2N-1}\big),
\end{equation}
contain in their highest order derivative term,
\begin{equation}\label{high_der}
\hat D(\nabla)\equiv\hat D^{a_1...a_{2N}}\nabla_{a_1}...\nabla_{a_{2N}},
\end{equation}
a nontrivial matrix $\hat D^{a_1...a_{2N}}$ contracting the indices of derivatives with the tensor indices of the field $\varphi^B(x)$ acted upon by $\hat H(\nabla)=H^A_B(\nabla)$. The simplest example of such operator is the vector field operator of the electromagnetic field in a generic $\alpha$-parameter gauge \cite{Barvinsky1985, GusyninGorbarRomankov, BarvinskyShapiro},
\begin{equation}\label{vector}
    D^a_b(\nabla)=-\Box\delta^a_b+\alpha\nabla^a\nabla_b,
\end{equation}
with $\alpha\neq 1$, or massive Proca model \cite{BarvinskyKalugin} with a particular value $\alpha=1$. More complicated examples are encountered, for example, in applications of Ho\v rava gravity model \cite{BarvinskyKurov2022}.

Various mathematical and physical aspects of heat kernel calculations for nonminimal operators were considered in numerous works \cite{Gusynin1991,GusyninGorbar,GusyninGorbarRomankov,GilkeyBransonFulling,BransonGilkeyPierzchalski,GuendelmanLeonidovNechitailoOwen1994,
AlexandrovVassilevich1996,AvramidiBranson2001,Avramidy2004,MossToms2014,IochumMason2017,GrassoKuzenko2023}, this list being very incomplete, but the systematic approach to this problem is likely to be largely missing. This is because known methods either do not preserve manifest diffeomorphism covariance, restrict calculations to lowest orders or the coincidence limit of the heat kernel instead of its off-diagonal structure, or do not clearly indicate the generic class of models for which the suggested technique is applicable.

The purpose of this paper is to extend the technique of heat kernel calculations to generic nonminimal operators (\ref{nonminimal_operator}) of {\em causal} theories, whose propagating modes in Lorentzian signature spacetime have characteristic surfaces associated with light cones in the metric $g_{ab}(x)$. For such models it is possible to express the local proper time expansion of their heat kernels in terms of the Schwinger--DeWitt coefficients of (\ref{DW}) with some auxiliary minimal operator of the form (\ref{minimal}). This technique of reduction from nonminimal to minimal operators, $e^{-\tau H}\to e^{-\tau F}$, will be very similar to the perturbation theory for the heat kernel, but it will be based not on the direct use of the two-point kernel (\ref{DW}) or its multiple convolutions in coordinate arguments. Rather it will be done at the level of noncommutative algebra of pseudodifferential operators, the final stage of it being a straightforward application of the Schwinger--DeWitt series (\ref{DW}).

To be more concrete, the needed exponential $e^{-\tau H}$ of the operator (\ref{nonminimal_operator}) will be linearly expressed by an integral operation acting on the minimal operator exponential $e^{-\tau' F}$.
\begin{equation}\label{operator_structure}
    e^{-\tau H(\nabla)}=\int d\mu(\tau')\,\mathfrak{B}(\tau,\tau',\mathfrak{R}\,|\,\nabla)\,e^{-\tau' F(\nabla)}
\end{equation}
This linear operation will involve multiple proper time integrations with a certain measure $d\mu(\tau')$ including the integration over $\tau'$, which would admit asymptotic expansion for $\tau\to 0$. The kernel of this linear map $\mathfrak{B}(\tau,\tau',\mathfrak{R}\,|\,\nabla)$ will be constructed via perturbation theory in powers of spacetime curvatures, background fields and their covariant derivatives, whose monomials of orders $r$ and $l$ we denote collectively by $\mathfrak{R}^r$ and $\nabla^l\mathfrak{R}^r$.\footnote{The operators (\ref{minimal}) and (\ref{nonminimal_operator}) are characterized by the set of curvatures $\mathfrak{R}=(R^a_{\;bcd},\hat{\cal R}_{ab},\hat P,...)$, including the Riemann tensor, fiber bundle curvature $\hat{\cal R}_{ab}=[\nabla_a,\nabla_b]$, the potential term $\hat P$ and other dimensional background fields contained in the coefficients of lower derivative terms of (\ref{nonminimal_operator}).} Each order of $\mathfrak{R}$ contributes to $\mathfrak{B}(\tau,\tau',\mathfrak{R}\,|\,\nabla)$ a term which is a {\em local differential operator} acting on $e^{-\tau' F(\nabla)}$ so that the generic term of expansion for $e^{-\tau H}\delta(x,x')$ has the form
\begin{equation}
    \nabla^l\mathfrak{R}^r(x)\nabla\nabla...\nabla\hat a_m(x,x') \label{terms-structure}
\end{equation}
with the coefficients given by multiple proper time integrals---the functions of the Synge world function $\sigma(x,x')$ and its multiple derivatives.

Important aspect of the coefficients of the terms (\ref{terms-structure}), given by multiple proper time integrals, is their finiteness. As we will see, the above formalism of reduction to minimal operators necessitates to consider the projectors to irreducible field components of the model, which in momentum representation involve negative powers of the momentum squared $p^2$. In the coordinate representation they become negative powers of $\Box$ or, more generally, negative powers of some minimal operator. Their representation in terms of the proper time integral, $(-\Box)^{-n}=\int_0^\infty d\tau\,\tau^{n-1} e^{\tau\Box}/(n-1)!$, involves infinite integration range. Under the term by term integration of the Schwinger--DeWitt series (\ref{DW}) for $e^{\tau\Box}$ this generically leads to infrared (IR) divergences at $\tau\to\infty$.

It is well-known that the actual power law decrease of $e^{\tau\Box}\delta(x,x')$ at $\tau\to\infty$, which allows one to solve this problem, can be attained within perturbation theory by a partial resummation of DeWitt series. This method, however, requires evisceration of Schwinger--DeWitt coefficients by decomposing them into the polynomials in curvatures $\mathfrak{R}$ and their covariant derivatives, $\hat a_m(x,x)\sim \sum_k\nabla^{2k}\mathfrak{R}^{m-2k}$, followed by the summation of infinite series of derivatives acting on a selected tensor structure of a given order in curvatures \cite{CPTII,Barvinsky2002,Barvinsky03}. This results in a nonlocal formfactor acting on a selected tensor term, but this is not what we want to do here, because our goal is the local expansion with the Schwinger-DeWitt coefficients kept intact within our formalism.

In fact, the problem of IR divergences should not arise at all, because in contrast to, say, the Green function $1/F$, its exponential is a well defined bounded (for positive $F$) entire function of $F$. So the coefficients of its asymptotic expansion in $\tau$ should be well defined and finite. This property, however, might be not manifest within a certain calculational scheme. So our goal will be to provide such a formalism in which no IR divergences arise at all stages of calculations. This will be attained by providing a finite range of integration over all proper time parameters in the coefficient functions of the terms (\ref{terms-structure}).

As we will see, the provision of a finite integration range is based on a certain type of subtraction procedure similar to the subtraction from a function of some of its Taylor expansion terms. We will develop two versions of this subtraction scheme. The simplest one allows one to reach the needed goal of eradicating IR divergences, but the subtracted (and then added back) terms form a set of differential, that is delta function like $\sim\nabla...\nabla\delta(x,x')$, contributions to the kernel $e^{-\tau H}\delta(x,x')$ while the rest comprises the integral term with a singular (though integrable) two-point kernel. This is a manifestation of the ultraviolet (UV) properties of the theory associated with the lower limit $\tau=0$ of integration over the proper time.

This observation is likely to contradict a common wisdom that the heat kernel is a smooth function of its two arguments because of a good bounded operator nature of the heat kernel. For this reason we also develop an alternative subtraction scheme, which allows one to remove the UV range of proper time integration at $\tau\to 0$, and thus demonstrate a smooth nature of the heat kernel. As a result, for $e^{-\tau H}\delta(x,x')$ we will have a good calculable coincidence limit at $x'=x$ in ${\rm Tr}\,e^{-\tau H}=\int d^dx\,e^{-\tau H}\delta(x,x')\,|_{\,x'=x}$, necessary for the calculation of the one-loop effective action. As the coincidence limits $\nabla^l\hat a_m(x,x')\,|_{\,x'=x}\propto\nabla^p\mathfrak{R}^r(x)$, $p+2r=2m+l$, are themselves local functions of curvatures and their derivatives, the series in terms of (\ref{terms-structure}), with coefficient functions finite at $\sigma(x,x)=0$, will represent a good local expansion in powers of background dimensionality \cite{Barvinsky1985}---a primary goal of effective field theory.

It turns out that this alternative subtraction scheme is unlikely to be possible for nonminimal operators with a degenerate (not invertible) principal symbol, which is the case of the Proca model. Such operators go beyond the conventional Gilkey--Seeley theory of differential operators \cite{Gilkey1975,Gilkey1995}, which at the quantum level in curved spacetime already in the one-loop approximation lead to double pole UV divergences \cite{Barvinsky1985,BarvinskyKalugin}. The heat kernel can be constructed for such models, but it is likely to be a distribution with singular two-point kernels. This will be illustrated for the Proca model operator as opposed to those with the nondegenerate principal symbol in the vector field case (\ref{vector}) with $\alpha\neq 1$, having a smooth heat kernel.

\section{Causal theories}
The class of causal theories can be defined by the condition that the matrix of the principal symbol of their wave operator (\ref{nonminimal_operator})---its highest order derivative term with the covariant derivatives replaced by momenta, $\nabla_a\to ip_a$,
\begin{equation}\label{symbol}
    \hat D(\nabla)\to\hat D(ip)=(-1)^N\hat D^{a_1...a_{2N}}p_{a_1}...p_{a_{2N}},
\end{equation}
has a determinant given by the power of momentum squared $p^2=g^{ab}p_ap_b$,
\begin{equation}\label{causality}
{\rm det}\,\hat D(ip)=C(p^2)^{NM},
 \end{equation}
with some momentum independent coefficient $C$. Causality of such theories is obvious because this property indicates that in Lorentzian signature spacetime the characteristic surfaces of field modes, defined by the equation ${\rm det}\,\hat D(ip)=0$ are null in the metric $g_{ab}(x)$. Similarly, the flat spacetime propagator in these theories, containing this determinant in the denominator, has poles corresponding to physical modes with $p^2=0$ (or $p^2+m_i^2=0$ if the matrix of masses $m_i$ is taken together with the principal symbol of the wave operator).

In fact, for Lorentz covariant theories the causality property implies that the matrix $\hat D^{a_1...a_{2N}}$, even though it is not diagonal in the space of fields, is built entirely of the metric, Kronecker symbols and Dirac matrices, and belongs to the representation of the Lorentz group. This makes ${\rm det}\,\hat D(ip)$ a function of a single scalar invariant $p^2$, and in the absence of other dimensional parameters it becomes a certain power of momentum squared.

Perturbation theory for the heat kernel begins with the approximation in which the operator is reduced to the contribution of its principal symbol, $\hat H(\nabla)\to\hat D(ip)$, retaining only its highest order term in Fourier momenta, $\hat D(p)\sim p^{2N}$. Then its heat kernel in the momentum space representation can be written down by using the eigenvalues and eigenvectors of this matrix-valued symbol. For causal theories this symbol is Lorentz covariant, and its eigenvalues---the functions of the only invariant $p^2=g^{ab}p_a p_b$---read as $\lambda_{i}(p^2)^N$ with some momentum independent coefficients $\lambda_{i}$. Then the heat kernel in this approximation equals
\begin{align} \label{A1}
&\hat K_H(\tau\,|\,x,y)\simeq\int \frac{d^dp}{(2\pi)^d}\,e^{ip(x-y)}\nonumber\\
&\qquad\qquad\times\sum_{i}e^{-\tau\lambda_{i}(p^2)^N}\hat\varPi_{i}(n), \;\;
n_a=\frac{p_a}{\sqrt{p^2}},
\end{align}
where $\hat\varPi_{i}(n)$ form the set of projectors on eigenspaces associated with eigenvalues of the principal symbol of the operator, which are the functions of the unit vector collinear with the momentum,
\begin{equation} \label{projectors}
\hat\varPi_{i}\hat\varPi_{k}=\delta_{ik}\hat\varPi_{i},\quad\sum_{i}\hat\varPi_{i}=\hat 1.
\end{equation}

These projectors for a completely symmetric bosonic spin $s$ field theory represent the matrices in the space of tensor labels $\hat\varPi_{i}(n)=\varPi_{(i)\,B}^{\;\;\;\;\,A}(n)$, whose multiindices, $A=a_1...a_s$, $B=b_1...b_s$, are composed of $s$ single spacetime indices. In view of covariance, the structure of the projector is the linear combination of monomials of unit vectors $n_a$ of a maximum power $2s$, because all $2s$ indices of the projector should be contracted with indices of momenta or the indices of the metric tensor.\footnote{For tensors with mixed symmetry this bound can be not saturated, but we will mainly consider the case of totally symmetric fields.} This can be summarized in the form of the expression
\begin{equation} \label{projectors1}
\hat\varPi_{i}=\hat\pi_{i}^{a_1...a_{2s}}\frac{p_{a_1}...p_{a_{2s}}}{(p^2)^s}
\end{equation}
with some momentum independent matrix coefficients. In the presence of fermions the structure is similar modulo additional gamma matrices factors which can be absorbed into $\hat\pi_{i}^{a_1...a_{2s}}$. The discussion of the linear algebra details of these projectors can be found in \cite{AvramidiBranson2001,Avramidy2004}.

\section{Leading order approximation\label{LOsection}}
Critical step of constructing a local {\em covariant} curvature expansion consists in building the leading order approximation which should be based not on flat spacetime objects, but on full covariant derivatives $\nabla_a$ and their covariant d'Alembertian $\Box=g^{ab}(x)\nabla_a\nabla_b$. It can be directly done by rewriting the formalism of Eqs. (\ref{A1})-(\ref{projectors1}) in the coordinate space and promoting all its objects to curved spacetime, $p_a\to-i\nabla_a$ and $p^2\to-\Box$. As a result these objects will satisfy the flat space relations modulo the corrections proportional to curvatures and other dimensionful background fields $\mathfrak{R}$, which we will denote as $O[\,\mathfrak{R}\,]$.

Under this operation the leading term of the heat kernel expansion takes the form
\begin{equation} \label{A2}
\hat K_H(\tau)\simeq\sum_{i}\hat\pi_{i}^{a_1...a_{2s}}\nabla_{a_1}...\nabla_{a_{2s}}\frac{e^{ -\tau\lambda_{i}(-\Box)^N}}{\Box^s}.
\end{equation}
Moreover, the $(-\Box)^N$-operator can be replaced with a more general minimal operator, $(-\Box)^N\to F=\hat F(\nabla)=(-\Box+\hat P)^N=(-\Box)^N+O[\,\mathfrak{R}\,]$ by including into it lower derivative terms with some background fields. This can be justified if it improves the properties of the curvature expansion. As we will see, for example, in the case of Proca operator in curved spacetime the inclusion of the Ricci term into the minimal operator $F^a_b(\nabla)$ can even make the leading order heat kernel expression exact, while without this term there would be extra nonlocal series in $R^a_b$.
From this point onward, we assume $N=1$. Note, that as soon as we do not explicitly substitute heat kernel expansions, this assumption goes without the loss of generality, as arbitrary $N$ can be restored through proper redefinitions.

Thus, we will define the following quasi-projectors
\begin{equation}\label{proj}
\varPi_{i}=\hat\varPi_{i}(\nabla)=\hat\pi_{i}^{a_1...a_{2s}}\nabla_{a_1}...\nabla_{a_{2s}}\frac{\hat 1}{[-\,F(\nabla)\,]^s},
\end{equation}
which now satisfy the relations (\ref{projectors}) only with $O[\,\mathfrak{R}\,]$ accuracy. In fact, the last of these relations, $\sum_{i}\varPi_{i}=1$, usually can be exactly enforced, because the highest spin irreducible projector is usually built from the requirement of this completeness relation.

\subsection{Subtraction procedure}
To avoid inevitable nonlocality in the set of these {quasi-}projectors (\ref{proj}), marred according to the discussion in Introduction by potential IR divergences, we can use the following subtraction procedure. Let the leading order heat kernel undergo the following sequence of identical transformations,
\begin{align}\label{subtraction}
K_H(\tau)&\simeq\sum_{i}\varPi_{i}e^{-\tau\lambda_{i}F}\nonumber\\
&=1+\sum_{i}\varPi_{i}\Big(e^{-\tau\lambda_{i}F}-1\Big)\nonumber\\
&=1-\sum_{i}\varPi_{i}F\lambda_{i}\int\limits_0^\tau d\tau_1\,e^{-\tau_1\lambda_{i}F}.
\end{align}
Repeating a similar subtraction procedure with the last exponential function and converting the result into a new integral in the finite range of the proper time we obtain
\begin{align}
K_H(\tau)&\simeq1-\tau\sum_{i}\varPi_{i}(F\lambda_{i})\nonumber\\
&\qquad-\sum_{i}\varPi_{i}F\lambda_{i}\int\limits_0^\tau d\tau_1\,
\Big(e^{-\tau_1\lambda_{i}F}-1\Big)\nonumber\\
&=1-\tau\sum_{i}\varPi_{i}(F\lambda_{i})\nonumber\\
&\qquad+\sum_{i}\varPi_{i}(F\lambda_{i})^2\int\limits_0^\tau d\tau_1
\int\limits_0^{\tau_1} d\tau_2\,e^{-\tau_2\lambda_{i}F}.
\end{align}
By continuing this procedure $n$ times we get
\begin{align}\label{main}
K_H(\tau)&\simeq\sum\limits_{k=0}^{n-1}\frac{(-\tau)^k}{k!}\sum_{i}\varPi_{i}(F\lambda_{i})^k\nonumber\\
&+(-1)^{n}\int\limits_0^\tau d^n\tau\,
\sum_{i}\varPi_{i}(F\lambda_{i})^n\,e^{-\tau_n\lambda_{i}F},
\end{align}
where
\begin{equation}\label{range}
\int\limits_0^\tau d^n\tau\,(...)\equiv\int\limits_0^\tau d\tau_1\int\limits_0^{\tau_1} d\tau_2...\int\limits_0^{\tau_{n-1}} d\tau_n\,(...).
\end{equation}
Thus, choosing $n\geq s$ we get in the second line of (\ref{main}) {\em local differential} operators of the form
\begin{equation}\label{PiF^n}
\varPi_{i}F^n=(-1)^s\hat\pi_{i}^{a_1...a_{2s}}\nabla_{a_1}...\nabla_{a_{2s}}\,\hat F^{n-s}(\nabla).
\end{equation}

The situation with the first line of (\ref{main}) looks more complicated, because it contains the operators of the so-called universal functional traces \cite{JackOsborn1984,Barvinsky1985} of the massless operator $F$, $\nabla...\nabla(1/F^{s-k})$, $k<s=n+1$, whose total symbol is singular at $p_a=0$. To generate them from the heat kernel $e^{-\tau F}$ one would need to integrate over an infinite range of the proper time $\tau$. As discussed in Introduction, this leads to IR divergent integrals at $\tau\to\infty$ if we work within Schwinger--DeWitt expansion with its coefficients staying intact.

However, there is an interesting observation that in flat-space approximation, that is with the original operator $H$ and the quasi-projectors $\varPi_i$ truncated to their principal symbol terms, the following identity holds for any power of $H$,
\begin{equation}\label{H^k}
H^k-\sum_{i}\varPi_{i}(F\lambda_{i})^k=O[\,\mathfrak{R}\,].
\end{equation}
This is, of course, the corollary of the eigenvalue problem for the principal symbol of $H$ in flat and empty spacetime. As the result, within the same zero-order accuracy in $\mathfrak{R}$ seemingly nonlocal coefficients belonging to the first line of (\ref{main}) can be replaced by the local powers of $H$, $\sum_{i}\varPi_{i}(F\lambda_{i})^k\to H^k$.

Thus, in the leading order approximation we can take as the heat kernel of $H$ the following modification of the expression (\ref{main}),
\begin{align}\label{LO}
\mathbb{K}_{n-1}(\tau)&=\sum\limits_{k=0}^{n-1}\frac{(-\tau)^k}{k!}H^k\nonumber\\
&+\int\limits_0^\tau d^n\tau\,
\sum_{i}\varPi_{i}(-F\lambda_{i})^n\,e^{-\tau_n\lambda_{i}F}.
\end{align}
In what follows we will denote the basic operators of the leading order approximation by double-struck letters.

The integer value $n\geq s$ will be chosen to provide locality of the operator factors in the needed curvature expansion. Note that the first $n$ terms of this expression represent the first $n$ orders of the Taylor expansion for $e^{-\tau H}$, while its $n$-fold multiple integral part is the subtracted version of the full heat kernel.\footnote{Obviously, we could have obtained \eqref{LO} in another way: firstly extract $n$ terms in Taylor expansion for operator exponential $e^{-\tau H}$, and only then rewrite the residual term using minimal operator $F$ and quasi-projectors $\varPi_i$.}

Perturbation theory in curvature on top of (\ref{LO}) is now straightforward and makes the solution of the heat kernel problem for causal operators completely systematic.

\section{Perturbation theory for the full heat kernel\label{Perturbation_section}}
As shown in Appendix \ref{A}, the nonvanishing right-hand side of the heat equation for the leading order heat kernel $\mathbb{K}_{n-1}(\tau)$,
\begin{equation}
\Big(\frac\partial{\partial\tau}+H\Big)\,\mathbb{K}_{n-1}(\tau)=-\mathbb{W}_{n-1}(\tau),
\end{equation}
equals
\begin{align}\label{Wn-1}
&\mathbb{W}_{n-1}(\tau)=(-1)^n\int\limits_0^\tau d^{n-1}\tau\,\bigg\{\Big[H^n-\sum_{i}H\,\varPi_{i}(F\lambda_{i})^{n-1}\Big]\nonumber\\
&\quad+\sum_{i}\Big[H\varPi_{i}(F\lambda_{i})^{n-1}-\varPi_{i}(F\lambda_{i})^n\Big]\,e^{-\tau_{n-1}\lambda_{i}F}\bigg\}
\end{align}
In view of the relation (\ref{H^k}) and the fact that all the commutators are vanishing for zero curvatures and zero values of dimensional background fields, $[H,\varPi_i]=O[\,\mathfrak{R}\,]$, $[F,\varPi_i]=O[\,\mathfrak{R}\,]$, this perturbation is at least linear in $\mathfrak{R}$, so that the perturbation theory in $\mathbb{W}_{n-1}$ will provide a needed curvature expansion,
\begin{equation}
\mathbb{W}_{n-1}=O[\,\mathfrak{R}\,].
\end{equation}

Moreover, by choosing $n=s+1$, which due to (\ref{PiF^n}) is the minimal value guaranteeing locality of both coefficients in square brackets of the above expression for $\mathbb{W}_{n-1}$, we get the perturbation $\mathbb{W}_s(\tau)$ free from negative powers of $F$,
\begin{align}\label{Ws}
&\mathbb{W}_s(\tau)=(-1)^{s+1}\int\limits_0^\tau d^s\tau\,\bigg\{\Big[H^{s+1}-\sum_{i}H\,\varPi_{i}(F\lambda_{i})^s\Big]\nonumber\\
&\quad+\sum_{i}\Big[H\varPi_{i}(F\lambda_{i})^s-\varPi_{i}(F\lambda_{i})^{s+1}\Big]\,e^{-\tau_s\lambda_{i}F}\bigg\}.
\end{align}
This completely saves us from the necessity of integration in the infinite proper time range---all integration ranges (\ref{range}) are bounded by the argument of $\mathbb{W}_s(\tau)$ and do not lead to IR divergences.

We look for the exact heat kernel $\mathbb{K}(\tau)$ of the nonminimal operator $H$, which satisfies the equation
\begin{equation}
\Big(\frac\partial{\partial\tau}+H\Big)\mathbb{K}(\tau)=0,
\end{equation}
in the form
\begin{equation}
\mathbb{K}(\tau)=\mathbb{K}_s(\tau)+\mathbb{V}_s(\tau),  \label{K_s+V}
\end{equation}
whence it follows that $\mathbb{V}_s(\tau)$ solves the initial value problem,
\begin{equation}
\Big(\frac\partial{\partial\tau}+H\Big)\mathbb{V}_s(\tau)=\mathbb{W}_s(\tau),\quad \mathbb{V}_s(0)=0.
\end{equation}
Its exact solution in terms of a yet unknown $\mathbb{K}$ is straightforward
\begin{equation}
\mathbb{V}_s(\tau)=\int\limits_0^\tau d\tau_1\,\mathbb{K}(\tau-\tau_1)\mathbb{W}_s(\tau_1),
\end{equation}
so that from (\ref{K_s+V}) we obtain the equation for $\mathbb{K}(\tau)$
\begin{equation}
\mathbb{K}(\tau)=\mathbb{K}_s(\tau)+\int\limits_0^\tau d\tau_1\,\mathbb{K}(\tau_1)\mathbb{W}_s(\tau-\tau_1),  \label{K=K_s+KV}
\end{equation}
easily solvable by iterations.
\begin{equation}
\mathbb{K}(\tau)=\mathbb{K}_s(\tau)+\sum\limits_{n=1}^\infty \mathbb{K}_s^{(n)}(\tau),  \label{full_series}
\end{equation}
where the curvature expansion terms, $\mathbb{K}_s^{(n)}(\tau)=O[\,\mathfrak{R}^n\,]$, read
\begin{align}
&\mathbb{K}_s^{(n)}(\tau)=\int\limits_0^\tau d^n\tau\,\mathbb{K}_s(\tau_n)\,\mathbb{W}_s(\tau-\tau_1)\nonumber\\
&\quad\times\mathbb{W}_s(\tau_1-\tau_2)\,\mathbb{W}_s(\tau_2-\tau_3)...
\mathbb{W}_s(\tau_{n-1}-\tau_n).  \label{K^n}
\end{align}

\subsection{Noncommutative algebra of pseudodifferential operators}
These expressions should be further rearranged as follows. To begin with, note that the structure of the leading order heat kernel (\ref{LO}) and the perturbation (\ref{Wn-1}) can be expressed as linear operations on the heat kernel of the minimal operator,
\begin{align}
&\mathbb{K}_s(\tau)=\int d\tau'\,K(\tau,\tau'\,|\,\nabla)\,e^{-\tau'F}, \label{K3}\\
&\mathbb{W}_s(\tau)=\int d\tau'\,W(\tau,\tau'\,|\,\nabla)\,e^{-\tau'F},\label{W3}
\end{align}
with some {\em local differential} operators $K(\tau,\tau'\,|\,\nabla)$ and $W(\tau,\tau'\,|\,\nabla)$. Substitution of these structures into (\ref{K^n}) leads to the representation very similar to the typical perturbation terms multilinear in the basic operator exponential $e^{-\tau F}$ and local ``vertices'' $W(\nabla)$,
\begin{align} \label{Kn1}
&\mathbb{K}_s^{(n)}(\tau)=\int\limits_0^\tau d^n\tau\,\int d\tau'_0\,K(\tau_n,\tau'_0|\,\nabla)\,e^{-\tau'_0F}\nonumber\\
&\quad\times\prod\limits_{m=1}^n \int d\tau'_m\,W(\tau_{m-1}-\tau_m,\tau'_m\,|\,\nabla)\,e^{-\tau'_m F}
\end{align}
However, instead of arranging the convolution of multiple kernels of $W(\nabla)e^{-\tau F}$ in the $x$-space of their arguments,
\begin{align}\label{convolution}
\cdots\int d^dx'\,WK_F(\tau\,|\,x,x')\,WK_F(\tau'\,|\,x',x'')\cdots,
\end{align}
we suggest, first, to commute all heat kernel operators to the right by using the following commutation relation
\begin{align}\label{commutator}
\big[\,e^{-\tau F},W\,\big]=\sum\limits_{n=1}^\infty\frac{(-\tau)^n}{n!}\overbrace{\big[F,[F,\cdots[F}^n,W]\cdots]]\,e^{-\tau F}
\end{align}
and noting that $\overbrace{\big[F,[F,\cdots[F}^n,W]\cdots]]=O\big[\,\nabla^n\mathfrak{R}\big]$ for a second-order operator $F$. This is because every commutator of a {\em minimal} operator with $W(\nabla)=O\big[\mathfrak{R}\big]$,
\begin{equation}
\big[F,W]=2(\nabla_aW)\nabla^a+(\Box W)+\ldots\propto\nabla\mathfrak{R}, \label{[FW]}
\end{equation}
increases the background dimensionality of the expression by one, that is adds at least one covariant derivative acting on the background fields contained in the coefficients of the differential operator $W(\nabla)$, while the analysis for higher-order minimal operator $F$ is similar.\footnote{Note that this mechanism of increasing background dimensionality works only for minimal operators with a unit matrix of the leading term, $F=(-\Box)^N\hat 1+\cdots$. Purely matrix commutators associated with noncommutativity of hatted matrices do not increase the dimensionality of differential operator coefficients.} Additional source of curvatures contributing to (\ref{[FW]}) $\mathfrak{R}^2$-type terms is the commutation of covariant derivatives, $[\nabla,\nabla]\sim\mathfrak{R}$.

This commutation procedure results in the relocation of all $e^{-\tau'_m F}$ to the right and the modification of the original differential operators $W(\tau,\tau'\,|\,\nabla)$ by these commutators of a higher background dimensionality, which are differential operators themselves. The overall heat kernel factor, acted upon from the left by differential operators of ever growing background dimensionality, thus takes the form $e^{-\tau' F}$ with $\tau'=\sum_{m=0}^n\tau'_m$. Therefore, the full series (\ref{full_series}) reads exactly as in Eq.(\ref{operator_structure}) with the operator coefficient $\mathfrak{B}(\tau,\tau',\mathfrak{R}\,|\,\nabla)$ given by a series in powers of the curvature $\mathfrak{R}$ and its derivatives. It consists of the monomials built of the above differential operators $K(\tau,\tau'\,|\,\nabla)$, $W(\tau,\tau'\,|\,\nabla)$ and commutator terms of (\ref{commutator}), along with the integration over all proper time parameters. This integration runs in a finite domain bounded from above by the time argument of $e^{-\tau H(\nabla)}$, which guarantees safety of the formalism in IR domain.

The last step of the calculational scheme consists in the substitution of the Schwinger--DeWitt expansion (\ref{DW}) into (\ref{operator_structure}), but before doing that we illustrate the above pseudodifferential calculus on the example of the Proca operator.

\section{Example of Proca field\label{Proca_section}}
Illustration of this technique is straightforward on the example of the vector Proca model with the nonminimal operator in curved spacetime
\begin{equation}
H^a_b(\nabla)=-\Box\delta^a_b+\nabla^a\nabla_b+R^a_b,  \label{Proca}
\end{equation}
where $R^a_b$ is the Ricci tensor. The choice of the relevant minimal operator is obvious
\begin{equation}
F^a_{\,b}(\nabla)=-\Box\delta^a_b+R^a_b.  \label{minimal_vector}
\end{equation}
It has remarkable properties---two identities understood as acting from the left respectively on a vector and a scalar,
\begin{equation}
\nabla_aF^a_{\,b}(\nabla)=-\Box\nabla_b,\quad F^a_{\,b}(\nabla)\nabla^b=-\nabla^a\Box,  \label{Ward}
\end{equation}
where $\Box$ is a scalar d'Alembertian. These identities allow one also to write down the following relations for the Green functions---the analogue of Ward identities \cite{Barvinsky1985}
\begin{equation}
\nabla_a\frac{\delta^a_{\,b}}{F(\nabla)}=-\frac1\Box\nabla_b,\quad \frac{\delta^a_{\,b}}{F(\nabla)}\nabla^b=-\nabla^a\frac1\Box.  \label{Ward1}
\end{equation}

The choice of the two projectors on subspaces of eigenmodes of the principal symbol of $H$ is also straightforward,
\begin{eqnarray}
&&\varPi_1=1+\nabla\nabla\frac1F\equiv\delta^a_{\,b}+\nabla^a\nabla_c\frac{\delta^c_{\,b}}{F(\nabla)}
=\delta^a_{\,b}-\nabla^a\frac1\Box\nabla_b,  \label{}\nonumber\\
&&\varPi_2=-\nabla\nabla\frac1F=\nabla^a\frac1\Box\nabla_b,
\end{eqnarray}
where we used an obvious notation $\nabla\nabla\equiv\nabla^a\nabla_b$. These projectors satisfy usual relations, $\varPi_1+\varPi_2=1$, $\varPi_1\varPi_2=0$, and their corresponding principal symbol eigenvalues read
\begin{equation}
\lambda_1=1,\quad \lambda_2=0,
\end{equation}
so that in index free notations
\begin{equation}
H=F+\nabla\nabla=\varPi_1F.
\end{equation}

For the vector case $s=1$ the leading order heat kernel (\ref{LO}) of $H$ reads in view of $\lambda_2=0$,
\begin{align}
\mathbb{K}_1(\tau)&=1-\tau H+\int\limits_0^\tau d^2\tau\,\varPi_1 F^2\,e^{-\tau_2 F}\nonumber\\
&=1-\tau F+\int\limits_0^\tau d^2\tau\,F^2\,e^{-\tau_2 F}\nonumber\\
&\quad+\int\limits_0^\tau d^2\tau\,\nabla\nabla F\,e^{-\tau_2F}-\tau (H-F),
\end{align}
where the first group of terms can be converted by integration over $d^2\tau$ back to the form of the heat kernel of the minimal operator $F$,
\begin{align}\label{LO1}
1-\tau H+\int\limits_0^\tau d^2\tau\,F^2\,e^{-\tau_2 F}=e^{-\tau F},
\end{align}
while in view of (\ref{Ward})
\begin{align}
&\int\limits_0^\tau d^2\tau\,\nabla\nabla F\,e^{-\tau_2 F}-\tau (H-F)\nonumber\\
&\quad=-\nabla\Box\int\limits_0^\tau d^2\tau\,e^{\tau_2\Box}\nabla-\tau\nabla\nabla
=\nabla\frac{1-e^{\tau\Box}}\Box\nabla.
\end{align}
Therefore, the leading order contribution to the heat kernel equals
\begin{align}
\mathbb{K}_1(\tau)&=e^{-\tau F}+\nabla\frac{1-e^{\tau\Box}}\Box\nabla\nonumber\\
&=[\,e^{-\tau F}\,]^{\,a}_{\,b}+\nabla^a\!\int\limits_0^\tau d\tau_1\,e^{\tau_1\Box}\nabla_b
\end{align}
and turns out to be the exact answer previously obtained in \cite{BarvinskyKalugin,Endo1984}. Exact nature of this result is compatible with the fact that the perturbation (\ref{Wn-1}) identically vanishes, $\mathbb{W}_1=0$, because both of its terms are zero,
\begin{align}
H^2-H\varPi_1F=0,\quad H\varPi_1F-\varPi_1F^2=0.
\end{align}
This example shows how drastically a calculation scheme can be simplified by a successful choice of a concrete minimal operator $F$.

\section{The problem of smoothness of the heat kernel}
After the operator algebra conversion of $e^{-\tau H}$ to the form (\ref{operator_structure}) linear in $e^{-\tau F}$ we can finally use the Schwinger--DeWitt expansion (\ref{DW}) and make the needed proper time integrations encoded in $\mathfrak{B}(\tau,\tau',\mathfrak{R}\,|\,\nabla)$. This will bring us to the coefficient functions of $\sigma(x,x')$ and its derivatives in front of the tensor structures (\ref{terms-structure}). We will consider the case of a second order minimal operator $F$. For higher-order minimal operators one should use instead of (\ref{DW}) the analogous expansion in terms of generalized exponential functions, built in \cite{Wach3,BKW2024}, whose application is also rather straightforward, though technically more cumbersome.

Bearing in mind the structure of the leading order heat kernel (\ref{LO}) with multi-fold proper time integrals
\begin{align}
\int\limits_0^\tau d^s\tau\,&e^{-\textstyle\frac{\sigma(x,x')}{2\lambda_i\tau_s}}\tau_s^{m-d/2}\nonumber\\
&=\int\limits_0^\tau d\tau'\,\frac{(\tau-\tau')^{s-1}}{(s-1)!}\,e^{-\textstyle\frac{\sigma(x,x')}{2\lambda_i\tau'}}\tau'^{m-d/2}
\end{align}
we get the set of integrals of the form,
\begin{align}
\int\limits_0^\tau d\tau'\,e^{-\textstyle\frac{\sigma(x,x')}{2\lambda_i\tau'}}\tau'^{\gamma-1}\propto\left[\frac{\sigma(x,x')}{2\lambda_i}\right]^\gamma,
\end{align}
which are singular at $x'\to x$ for several negative values of $\gamma=m+1-d/2$ corresponding to low numbers $m$ of Schwinger-Dewitt coefficients. These are the terms associated with relevant and marginal operators of effective field theory, responsible for ultraviolet renormalization.

Moreover, the presence of local differential operators in the leading order approximation---the first  line of (\ref{LO})---implies that there will also be such distribution terms $\sim\nabla\nabla\cdots\nabla\delta(x,x')$ in $\mathfrak{B}(\tau,\tau',\mathfrak{R}\,|\,\nabla)$ not containing the factor of $e^{-\tau F}$ at all. This obviously contradicts the general consideration that the heat kernel of elliptic operator in view of its boundedness should be a smooth function of its spacetime arguments.

The whole answer for a heat kernel of the nonminimal $H(\nabla)$ might be still a smooth object in view of certain cancellations of singular, but integrable, parts, so that these singularities might be an artifact of a somewhat inappropriate subtraction and perturbation scheme that we used.\footnote{The fact that the coordinate kernel of a well-defined operator $\nabla\nabla(1/\Box)$ consists of an integral part, supplied with a special integration prescription, plus a certain local term is known to underlie the cancellation of volume UV divergences in the one-loop effective action in curved spacetime and rules out the necessity of cosmological constant renormalization in massless theories \cite{Fradkin1977, TwoLoop1987}.} What is responsible for the origin of these singularities is in fact integration over the proper time parameter in the vicinity of zero. This is the UV domain of energy scales corresponding to the short distance limit $\sigma(x,x')\to 0$. This observation suggests the way to circumvent this problem by raising the lower integration limit for $\tau$ from zero. This is achieved by the following modification of the subtraction procedure and the perturbation scheme.

\section{Alternative subtraction procedure and perturbation scheme \label{Alternative}}
To shift the lower $\tau$-integration limit from zero, one should replace the subtraction of local (unity) operator in (\ref{subtraction}) by the heat kernel of $F$ at some nonzero value of the proper time. Let us take this value to be $\beta\tau$ with some positive parameter $\beta$, $1\to e^{-\beta\tau F}$, and represent the arising difference as the $\tau$-integral,
\begin{align}\label{beta_subtraction}
&e^{-\beta\tau F}+\big(e^{-\tau\lambda_{i}F}-e^{-\beta\tau F}\big)\nonumber\\
&\qquad\qquad\qquad=e^{-\beta\tau F}-F\int\limits_{\beta\tau}^{\lambda_i\tau} d\tau_1\,e^{-\tau_1 F},
\end{align}
Repeating this subtraction $s+1$ times and making a similar replacement, just like in Sect.\ref{LOsection}, $\sum_{i}\varPi_{i}(F\lambda_{i})^k\to H^k$, one gets the new leading order approximation,
\begin{align}\label{LObeta}
\mathbb{K}^{(\beta)}_s(\tau)&=\sum\limits_{k=0}^s\frac{\tau^k}{k!}(\beta F-H)^k\,e^{-\beta\tau F}\nonumber\\
&+\sum_{i}\int\limits^{\lambda_i\tau}_{\beta\tau} d^{s+1}\tau\,
\varPi_{i}(-F)^{s+1}\,e^{-\tau_{s+1} F},
\end{align}
where the new definition of multi-fold integration measure and its relation to the old one (\ref{range}) read as
\begin{align}\label{range1}
\int\limits^{\lambda\tau}_{\beta\tau} d^n\tau\,f(\tau_1,&...\tau_n)\equiv\int\limits_{\beta\tau}^{\lambda\tau} d\tau_1\int\limits_{\beta\tau}^{\tau_1} d\tau_2...\int\limits_{\beta\tau}^{\tau_{n-1}} d\tau_n\,f(\tau_1,...\tau_n)\nonumber\\
&=\int\limits^{(\lambda-\beta)\tau}_0 d^n\tau\,f(\tau_1+\beta\tau,...\tau_n+\beta\tau).
\end{align}
Note that for $n>1$, $\int^a_b d^n\tau(\cdots)\neq -\int_a^b d^n\tau(\cdots)$, the equality holding only for $n=1$.

All the terms in (\ref{LObeta}) contain the heat kernel factors, either $e^{-\beta\tau F}$ or $e^{-\tau_{s+1} F}$ with $\tau_{s+1}$ integrated over. Thus, this leading order kernel does not contain purely differential local terms, which testifies in favor of its smooth properties. It is easy to check that this expression at $\beta=0$ with replaced integration parameters $\tau_m\to\lambda_i\tau_m$, $m=1,\ldots s$, goes over into (\ref{LO}). It is also straightforward to see that in the flat-space approximation the modified kernel is independent of $\beta$, because its $\beta$-derivative reduces to commutators proportional to the curvature,
\begin{align}\label{LObeta_dependence}
\frac\partial{\partial\beta}\mathbb{K}^{(\beta)}_s(\tau)=O[\,\mathfrak{R}\,].
\end{align}
Therefore, this $\beta$-modification effects only the perturbation theory terms.

Perturbation theory built on top of the leading order (\ref{LObeta}) repeats the formalism of Sect.\ref{Perturbation_section}. Leading order heat kernel satisfies the equation
\begin{align}\label{Wbeta}
&\Big(\frac\partial{\partial\tau}+H\Big)\,\mathbb{K}_s^{(\beta)}(\tau)=-\mathbb{W}_s^{(\beta)}(\tau),
\end{align}
where the perturbation $\mathbb{W}_s^{(\beta)}(\tau)$, as shown in Appendix~\ref{A}, equals
\begin{align}\label{Wsbeta}
\mathbb{W}_s^{(\beta)}(\tau)=&\sum\limits_{{k=1}}^{s}\frac{\tau^k}{k!}\big[\,(\beta F-H)^k,\beta F\,\big]\, e^{-\beta\tau F}\nonumber\\
&-\sum_{i}\int\limits^{\lambda_i\tau}_{\beta\tau} d^s\tau\,\Big[H\varPi_{i}-\varPi_{i}\lambda_{i}F\Big](-F)^s\,e^{-\tau_s F}\nonumber\\
&+\frac{\tau^s}{s!}\bigg[-\sum_{i}\big[\,\beta\varPi_{i}F-H\varPi_{i}\big]\big[\beta F-F\lambda_{i}\big]^{s}\nonumber\\
&\qquad\;\;\;+(\beta F-H)^{s+1}\,\bigg]\,e^{-\beta\tau F}.
\end{align}
It is more complicated than the expression (\ref{Wn-1}) and reduces to it at $\beta=0$. However, the freedom to choose the parameter $\beta$ allows one to simplify it---in particular, choosing $\beta$ to coincide with one of the eigenvalues $\lambda_i$ removes the corresponding integral term from (\ref{Wsbeta}).

Next orders of the expansion, in analogy with (\ref{full_series}) and (\ref{K^n}), obviously generalize to the expressions
\begin{align}
&\mathbb{K}^{(\beta)}(\tau)=\mathbb{K}^{(\beta)}_s(\tau)+\sum\limits_{n=1}^\infty \mathbb{K}_s^{(\beta,n)}(\tau),\nonumber\\
&\mathbb{K}_s^{(\beta,n)}(\tau)=\int\limits_0^\tau d^n\tau\,\mathbb{K}^{(\beta)}_s(\tau_n)\,\mathbb{W}^{(\beta)}_s(\tau-\tau_1)\nonumber\\
&\times\mathbb{W}^{(\beta)}_s(\tau_1-\tau_2)\,\mathbb{W}^{(\beta)}_s(\tau_2-\tau_3)...
\mathbb{W}^{(\beta)}_s(\tau_{n-1}-\tau_n),
\label{Ksbeta}
\end{align}
followed by the commutator calculus of Sect.\ref{Perturbation_section}. However, the integration range in these expressions runs from zero, but this should not induce local singular terms because neither $\mathbb{K}^{(\beta)}_s(\tau)$ nor $\mathbb{W}^{(\beta)}_s(\tau)$ contain them. Anticipated absence of singular integral terms with negative powers of $\sigma(x,x')$ will be checked below on the concrete example of the nonminimal vector field operator with the nondegenerate symbol.

It is clear, however, that there is a class of models in which the $\beta$-modification of the above type does not work. This is the case when one of the eigenvalues of the principal symbol of $H(\nabla)$ is zero. In this case the corresponding integration limit in (\ref{LObeta}) is zero, $\lambda_i\tau=0$, and the removal of the UV integration domain is impossible. But this is the case of a degenerate symbol of the nonminimal operator, when the standard Gilkey--Seeley formalism does not directly apply. This case bears a number of peculiarities, partly discussed in \cite{BarvinskyKalugin}, among which is the distributional nature of the heat kernel in coordinate space.

\section{Nonminimal vector field operator with a nondegenerate symbol\label{Vector_section}}
The application of modified subtraction procedure to the nonminimal vector field operator
\begin{equation}
H^a_b(\nabla)=-\Box\delta^a_b+\alpha\nabla^a\nabla_b+R^a_b,  \label{nonminimal-vector}
\end{equation}
with the nondegenerate symbol, $\alpha\neq 1$, is straightforward. Choosing $\beta=1$ {(this corresponds to one of the eigenvalues of the principal symbol)} and the minimal operator $F$ given by (\ref{minimal_vector}) we immediately get the leading order heat kernel,
\begin{align}\label{K11}
\mathbb{K}^{(1)}_1(\tau)&=e^{-\tau F}+\nabla\frac{e^{(1-\alpha)\Box}-e^{\tau\Box}}\Box\nabla\nonumber\\
&=e^{-\tau F}+\!\!\!\!\int\limits_\tau^{(1-\alpha)\tau}\!\! d\tau_1\,\nabla e^{\tau_1\Box}\nabla,
\end{align}
which turns out to be exact, just like in the degenerate case of the Proca model. The corresponding perturbation (\ref{Wsbeta}) is also zero in view of remarkable exact relations (\ref{Ward}) for $F=F^a_b(\nabla)$.

Thus the exact two-point heat kernel takes the form
\begin{align}\label{fullKforvec}
\big[\,K_H\big]^{\,a}_{\,b'}&(\tau\,|\,x,x')=\big[\,K_F\big]^{\,a}_{\,b'}(\tau\,|\,x,x')\nonumber\\
&-\nabla^a\nabla_{b'}\!\!\!\!\int\limits_\tau^{(1-\alpha)\tau}\!\!\! d\tau_1\,K_{-\Box}(\tau\,|\,x,x'),
\end{align}
where the primed index of the covariant derivative indicates that it is acting on the argument $x'$ of the kernel. The equivalent form of this result was previously obtained in~\cite{Endo1984}.

Using the expansion (\ref{DW}) in the second integral term and integrating in the finite range of the proper time, we have the series in Schwinger-DeWitt coefficients of the scalar operator $-\Box$,
\begin{align}\label{nonminimal_term_1}
&\int\limits_\tau^{(1-\alpha)\tau}\!\! d\tau_1\,K_{-\Box}(\tau\,|\,x,x')
=\frac{\Delta^{1/2}(x,x')}{(4\pi)^{d/2}}\, g^{1/2}(x')\nonumber\\
&\times\sum\limits_{m=0}^\infty I\big(\tau,m-\tfrac{d}2+1,\a\,\big|\,\sigma(x,x')\big)\,a_m(-\Box\,|\,x,x').
\end{align}
The coefficient function of $\sigma(x,x')$ is given here in terms of the upper incomplete gamma function, $\Gamma(p,x)=\int_x^\infty dy\,y^{p-1}e^{-y}$,
\begin{align}\label{Igamma}
&I\big(\tau,\gamma,\a\,\big|\,\sigma\big)=\!\!\!\!\int\limits_\tau^{(1-\alpha)\tau}\!\!\! d\tau_1\,
e^{\textstyle{-\frac{\sigma}{2\tau_1}}}\,\tau_1^{\gamma-1}\nonumber\\
&\quad=\Big(\frac\sigma2\Big)^\gamma\Big[\,\Gamma\Big(\!\!-\gamma,\frac\sigma{2\tau(1-\alpha)}\Big)-
\Gamma\Big(\!\!-\gamma,\frac\sigma{2\tau}\Big)\,\Big].
\end{align}

Its expansion in $\sigma$, which can be obtained via the term by term integration of the Taylor expansion of the integrand, reads
\begin{align}\label{Iexpansion}
&I\big(\tau,\gamma,\a\,\big|\,\sigma\big)=\tau^\gamma\sum\limits_{n=0}^\infty\Big(\!\!-\frac\sigma{2\tau}\Big)^n\frac{1-(1-\alpha)^{\gamma-n}}{n!(n-\gamma)},
\end{align}
where for positive integer $\gamma=k$, which is the physically interesting case of even spacetime dimension, the term at $n=k$ becomes logarithmic in $(1-\alpha)$-parameter, $(-\sigma/2)^k\ln(1-\alpha)/k!$. Thus it is regular in the coincidence limit $\sigma(x,x')=0$ as expected for operators with a nondegenerate principal symbol.

With two derivatives acting on (\ref{nonminimal_term_1}) in (\ref{fullKforvec}) one can write down the total heat kernel of the nonminimal vector operator $H(\nabla)$ in the nondegenerate case. In particular, in view of the coincidence limits for $I\big(\tau,\gamma,\a\,\big|\,\sigma\big)$
\begin{align}
&I(\tau,\gamma,\a\,|\,0)=-\tau^\gamma\frac{1-(1-\alpha)^\gamma}\gamma,\nonumber\\
&\nabla^a I(\tau,\gamma,\a\,|\,\sigma(x,x'))\,\big|_{\,x'=x}=0\nonumber\\
&\nabla^a\nabla_{b'}I(\tau,\gamma,\a\,|\,\sigma(x,x'))\,\big|_{\,x'=x}=-\tau^{\gamma-1}\frac{\delta^a_b}{2}\frac{1-(1-\alpha)^{\gamma-1}}{\gamma-1},\nonumber
\end{align}
one finds that the total coincidence limit $[\,K_H]^{\,a}_{\,b}(\tau\,|\,x,x)$ expresses in terms of the DeWitt coefficients of the minimal operator $F$ and those of the scalar d'Alembertian $(-\Box)$ along with their second order derivatives at $x'=x$,
\begin{align}
&\big[\,K_H\big]^{\,a}_{\,b}(\tau\,|\,x,x)
=\frac{g^{1/2}(x)}{(4\pi\tau)^{d/2}}\sum\limits_{m=0}^\infty \tau^m\bigg\{\,\big[\,a_m(F\,|\,x,x)\,\big]^{\,a}_{\,b}\nonumber\\
&\quad+\bigg[\frac{1-(1-\alpha)^{m-\frac{d}2}}{m-\frac{d}2}\frac{\delta^a_b}2+\frac{1-(1-\alpha)^{m-\frac{d}2+1}}{m-\frac{d}2+1}\nonumber\\
&\qquad\times\tau\Big(-\frac16 R^a_b+\nabla^a\nabla_{b'}\Big)\,\bigg]\,a_m(-\Box\,|\,x,x')\,\Big|_{\,x'=x}\bigg\}.
\end{align}
This result matches with that of \cite{BarvinskyShapiro} obtained for low $m$.

For $\alpha\to 1$ the expansion (\ref{Iexpansion}) is not applicable, and in this limit the first incomplete gamma function in (\ref{Igamma}) vanishes, while the second one generates in (\ref{nonminimal_term_1}), for UV sensitive part of the expansion, $\gamma=m+1-d/2<0$, the contribution singular at $x'\to x$,
\begin{equation}
I\big(\tau,\gamma,1\,\big|\,\sigma\big)=-\Big(\frac\sigma2\Big)^{m+1-d/2}
\Gamma\Big(\frac{d}2-1-m,\frac\sigma{2\tau}\Big).
\end{equation}
This confirms our anticipations of coincidence limit singularities in the heat kernel of operators with a degenerate symbol.

In view of asymptotic behavior of incomplete gamma functions, the large $\sigma$ limit of the function (\ref{Igamma}) reads
\begin{align} \label{IncomplAsympt}
&I\big(\tau,\gamma,\a\,\big|\,\sigma\big)=\frac2\sigma\sum\limits_{k=0}^\infty \frac{\Gamma(-\gamma)}{\Gamma(-\gamma-k)}\Big(\frac2\sigma\Big)^k\tau^{k+\gamma+1}\nonumber\\
&\quad\times\Big[\,(1-\alpha)^{k+\gamma+1}e^{-\tfrac\sigma{2\tau(1-\alpha)}}-
e^{-\tfrac\sigma{2\tau}}\Big] ,\;\sigma\to\infty.
\end{align}
Thus it reproduces the exponential type decrease of the minimal operator $F$ contribution~(\ref{DW}), strengthened by extra power law falloff.

\subsection{The case of a generic potential term}
Here we illustrate our calculational method for the case when the perturbation theory becomes nontrivial and higher order terms of the expansion (\ref{Ksbeta}) should be taken into account. As an example consider the nonminimal vector field operator with a generic potential term which does not reduce to the Ricci tensor,
\begin{equation}
H^a_b(\nabla)=-\Box\delta^a_b+\alpha\nabla^a\nabla_b+R^a_b+P^a_b.  \label{generic_potential}
\end{equation}
The set of curvatures characterizing this operator $\mathfrak{R}=(P^a_b,R^a_{bcd},R^a_b)$ now includes a generic correction $P^a_b$ to the full potential term. For simplicity of illustration we will obtain the coincidence limit of its heat kernel in the lowest nontrivial order of approximation---the second order in background dimensionality $O\big[\tfrac1{l^2}\big]$. As the potential itself  saturates this dimensionality $P^a_b=O\big[\tfrac1{l^2}\big]$, this means that we have to take into account all correction terms linear in the {\em undifferentiated} $P^a_b$.

With the same choice of $\beta=1$ the expressions (\ref{LObeta}) and (\ref{Wsbeta}) now take the form
\begin{align}
&\mathbb{K}^{(1)}_1(\tau)=(1-\tau P)\,e^{-\tau F}+\!\!\!\!
\int\limits_\tau^{(1-\alpha)\tau}\!\! d\tau'_0\,\nabla\nabla e^{-\tau'_0 F},\label{K11P}\\
&\mathbb{W}^{(1)}_1(\tau)=\tau(\,\alpha\nabla\nabla P+P^2+[F,P]\,)\,e^{-\tau F}\nonumber\\
&\qquad\qquad\qquad\qquad\qquad-P\!\!\!\!\int\limits_\tau^{(1-\alpha)\tau}\!\! d\tau'_1\,\nabla\nabla e^{-\tau'_1 F},\label{W11P}
\end{align}
where $\nabla\nabla$ denotes the matrix valued operator $(\nabla\nabla)^a_b=\nabla^a\nabla_b$ with all the derivatives acting upon everything to the right, so that $\nabla\nabla P$ should read
\begin{align}
(\nabla\nabla P)^a_b=&P^c_b\nabla^a\nabla_c+(\nabla^a P^c_b)\nabla_c+(\nabla_c P^c_b)\nabla^a\nonumber\\
&+(\nabla^a\nabla_c P^c_b)=P^c_b\nabla^a\nabla_c+O\Big[\frac1{l^3}\Big].
\end{align}

Comparing (\ref{K11}) with (\ref{K11P}) and noting that $\mathbb{W}^{(1)}_1(\tau)\propto P=O\big[\tfrac1{l^2}\big]$ we see that the contribution of $P$ to the heat kernel in this approximation reads
\begin{align}
K_H(\tau)&-K_H(\tau)\,\big|_{\,P=0}=-\tau P\,e^{-\tau F}\nonumber\\
&+\int\limits_0^\tau d\tau_1\,\mathbb{K}^{(1)}_1(\tau_1)\,\mathbb{W}^{(1)}_1(\tau-\tau_1)
+O\Big[\frac1{l^3}\Big], \label{}
\end{align}
where the first term gives the contribution to the coincidence limit
\begin{align}\label{first}
-\big[\,\tau P\,e^{-\tau F}\big]^{\,a}_{\,b'}\,\Big|_{\,x'=x}=-\frac{g^{1/2}(x)}{(4\pi\tau)^{d/2}}\tau P^a_b+O\Big[\frac1{l^3}\Big].
\end{align}
The perturbation can be truncated to
\begin{align}
&\big[\,\mathbb{W}^{(1)}_1(\tau)\big]^{\,a}_{\,b'}=\tau\,\alpha\, P^c_d\,\nabla^a\nabla_c\big[\,e^{-\tau F}\big]^{\,d}_{\,b'}\nonumber\\
&\qquad\quad -P^a_c\!\!\!\!\int\limits_\tau^{(1-\alpha)\tau}\!\! d\tau'_1\,\nabla^c\nabla_d \big[\,e^{-\tau'_1 F}\big]^{\,d}_{\,b'}
+O\Big[\frac1{l^3}\Big],
\end{align}
Using (\ref{K11P}) and freely moving all operator exponents to the right along with the freely commuted covariant derivatives---all relevant non-vanishing commutators go beyond our approximation---we obtain
\begin{widetext}
\begin{align}\label{second}
\Big[\int_0^\tau d\tau_1\,\mathbb{K}^{(1)}_1(\tau_1)\,\mathbb{W}^{(1)}_1(\tau-\tau_1)\Big]^{\,a}_{\,b'}&=\alpha\int_0^\tau d\tau_1\,
(\tau-\tau_1) P^c_d\,\nabla^a\nabla_c\big[\,e^{-\tau F}\big]^{\,d}_{\,b'}-\int_0^\tau d\tau_1\,
P^a_c \int_{\Delta\tau}^{(1-\alpha)\Delta\tau}\!\!\!\!d\tau'_1\,\nabla^c\nabla_d \big[\,e^{-(\tau_1+\tau'_1) F}\big]^{\,d}_{\,b'}\nonumber\\
&+\alpha\int_0^\tau d\tau_1\,(\tau-\tau_1) \int_{\tau_1}^{(1-\alpha)\tau_1}\!\!\!\!d\tau'_0\,P^d_e\,\nabla^a\nabla_c
\nabla^c\nabla_d\big[\,e^{-(\tau-\tau_1+\tau'_0) F}\big]^{\,e}_{\,b'}\nonumber\\
&-\int_0^\tau d\tau_1\,\int_{\tau_1}^{(1-\alpha)\tau_1}\!\!\!\!d\tau'_0
\int_{\Delta\tau}^{(1-\alpha)\Delta\tau}\!\!\!\!d\tau'_1\,P^c_d\nabla^a\nabla_c\nabla^d\nabla_e \big[\,e^{-(\tau'_1+\tau'_0) F}\big]^{\,e}_{\,b'}
+O\Big[\frac1{l^3}\Big],
\end{align}
where $\Delta\tau=\tau-\tau_1$.
\end{widetext}

Off-diagonal expression even in this lowest nontrivial order looks very complicated in view of numerous terms with derivatives $\nabla_a\sigma(x,x')=\sigma_a(x,x')\neq 0$ not to say about multiple proper time integrals which, however, reduce by integration by parts to linear combinations of (\ref{Igamma}) with various values of $\gamma$. For example,
\begin{align}\label{sample_integral}
&\int_0^{\tau}d\tau_1\int_{\Delta\tau}^{(1-\a)\Delta\tau}\!\!\!\!\!d\tau_1'(\tau_1+\tau_1')^{\g-1}\,
e^{-\frac{\s}{2(\tau_1+\tau_1')}}\nonumber\\
&\qquad
=\frac1\alpha I(\tau,\gamma+1,\alpha\,|\,\s)+\frac{\alpha-1}\alpha\tau I(\tau,\gamma,\alpha\,|\,\s).
\end{align}

The reduction to heat kernel diagonal simplifies the answer---integrals become elementary and the non-vanishing coincidence limits of the covariant
derivatives of $\sigma(x,x')$ in
\begin{align}
&\nabla_{a_1}\cdots\nabla_{a_n}\big[\,e^{-\tau F}\delta(x,x')\,\big]^{\,a}_{\,b}\,\Big|_{\,x'=x}\nonumber\\
&\quad=\frac{g^{1/2}(x)\delta^a_b}{(4\pi\tau)^{d/2}}\nabla_{a_1}\cdots\nabla_{a_n}
e^{\textstyle{-\frac{\sigma(x,x')}{2\tau}}}\,\Big|_{\,x'=x}\!\!+O\Big[\frac1{l}\Big]
\end{align}
reduce to $\nabla_a\nabla_b\sigma(x,x')\,|_{x'=x}=g_{ab}$. Thus, after combining (\ref{first}) and (\ref{second})
we finally obtain
\begin{align}
&\Big[\,K_H(\tau\,|\,x,x)-K_H(\tau\,|\,x,x)\,\big|_{\,P=0}\,\Big]^{\,a}_{\,b}\nonumber\\
&\qquad=\frac{g^{1/2}(x)}{(4\pi \tau)^{d/2}}\,\tau\,
\bigg\{\,\bigg[\,\frac{3\alpha(d+2)-4d-4}{\alpha d(d^2-4)(1-\alpha)^{d/2}}\nonumber\\
&\qquad\qquad\qquad\qquad\quad+\frac{\alpha(3- d^2)(d-2)+4 d+4}{\alpha  d \left(d^2-4\right)}\bigg]
\,P^a_b\nonumber\\
&\qquad\qquad+\bigg[\frac{-\alpha(d+2)+4}{\alpha  d(d^2-4)(1-\alpha )^{d/2}}
-\frac{\alpha(d-2)+4}{\alpha  d(d^2-4)}\bigg]\nonumber\\
&\qquad\qquad\qquad\times
(g^{ad}P^c_d\,g_{bc}+\d^a_b P^c_c)\bigg\}+O\lb\frac{1}{l^3}\rb.
\end{align}

\section{General term of perturbation theory\label{B}}
The results of the previous section allow us to build a generic term of the expansion (\ref{Ksbeta}) in the form of explicit proper time integrals.
Eqs.(\ref{K3})-(\ref{W3}) following from the expressions for $\mathbb{K}{}^{(\beta)}_s(\tau)$ and $\mathbb{W}{}^{(\beta)}_s(\tau)$ contain in the right hand sides both local and integral terms in the proper time, the limits of integration being dependent on eigenvalues $\lambda_i$. Therefore the precise structure of these equations can be rendered the form
\begin{align}
&\mathbb{K}^{(\beta)}_s(\tau)=\sum\limits_i\int\limits_{\beta\tau}^{\lambda_i\tau} d\tau'\,K_i(\tau,\tau')\,e^{-\tau'F}, \label{K4}\\
&\mathbb{W}^{(\beta)}_s(\tau)=\sum\limits_i\int\limits_{\beta\tau}^{\lambda_i\tau}  d\tau'\,W_i(\tau,\tau')\,e^{-\tau'F},\label{W4}
\end{align}
if we construct the {\em curved spacetime} projectors exactly satisfying completeness relation
\begin{align}
\sum_i\varPi_i\equiv\sum_i\varPi_i(\nabla)=1.
\end{align}
This, in fact, can always be done because as usual the main projector on irreducible field components is determined from the completeness condition.

With this assumption it follows from Eqs.(\ref{LObeta}) and (\ref{Wsbeta}) that the operators $K_i(\tau,\tau')$ and $W_i(\tau,\tau')$ have the form
\begin{align}\label{LOK}
K_i(\tau,\tau')&=\sum\limits_{k=0}^s\frac{\tau^k}{k!}\varPi_i(\beta F-H)^k\,\delta(\tau'-\beta\tau)\nonumber\\
&\quad-\frac{(\tau'-\lambda_i\tau)^s}{s!}
\varPi_{i}F^{s+1}
\end{align}
and
\begin{align}\label{W5}
W_i(\tau,&\tau')=\bigg\{\,\sum\limits_{k=0}^{s}\frac{\tau^k}{k!}\varPi_i\big[\,(\beta F-H)^k,\beta F\,\big]\nonumber\\
&-\frac{\tau^s}{s!}\Big((\,\beta\varPi_{i}F-H\varPi_{i}) F^s(\beta-\lambda_{i})^{s}\nonumber\\
&\qquad\quad-\varPi_i(\beta F-H)^{s+1}\Big)\,\bigg\}\,\delta(\tau'-\beta\tau)\nonumber\\
&+\frac{(\tau'-\lambda_i\tau)^{s-1}}{(s-1)!}(H\varPi_{i}-\lambda_{i}\varPi_{i}F)F^s\,.
\end{align}
They contain delta-function type terms in the proper time with extra projector factor $\varPi_i$-dependence without the nonlocality-compensating factor $F^s$. But this does nor ruin the local expansion formalism, because under the summation over $i$ nonlocal structures cancel out due to $\sum_i\varPi_i=1$ (note that in these non-integral terms the limits of integration $\lambda_i\tau$ are not involved).

Using (\ref{K4}) and (\ref{W4}) in (\ref{Ksbeta}) for the full set of the expansion orders
we get
\begin{align}\label{W6}
&\mathbb{K}_s^{(\beta,n)}(\tau)
=\int\limits_0^\tau d^n\tau\!\!\sum\limits_{i_0,i_1,\ldots i_n}\int\limits_{\beta\tau_n}^{\lambda_{i_0}\tau_n}d\tau'_0
K_{i_0}(\tau_n,\tau'_0)\,
\nonumber\\
&\quad\times
\Bigg\{\prod\limits_{m=1}^n e^{-\tau'_{m-1}F}\!\!\!
\int\limits_{\beta\Delta\tau_m}^{\lambda_{i_m}\Delta\tau_m}\!\!\!d\tau'_m
W_{i_m}(\Delta\tau_m,\tau'_m)\Bigg\}\,e^{-\tau'_n F},
\end{align}
where $\Delta\tau_m\equiv\tau_{m-1}-\tau_m$ and $\tau_0\equiv\tau$.

Denoting for brevity $W_{i_m}(\tau_{m-1}-\tau_m,\tau'_m)=W_{i_m}$, one can rewrite the expression from curly brackets of (\ref{W6}) as
\begin{align}\label{comm0}
&\prod\limits_{m=1}^n e^{-\tau'_{m-1}F}\,W_{i_m}=\bigg[\prod\limits_{m=1}^{n-1} e^{-\tau'_{m-1}F}\,W_{i_m}\!\bigg]\,e^{-\tau'_{n-1}F}\,W_{i_n},
\end{align}
where the factor outside of square brackets can be transformed by commuting the operator exponent to the right
\begin{align}
&e^{-\tau'_{n-1}F}\,W_{i_n}=\widetilde W_{i_n}e^{-\tau'_{n-1}F},\label{comm}\\
&\widetilde W_{i_n}=\sum\limits_{k_n=0}^\infty\frac{(-\tau'_{n-1})^{k_n}}{k_n!}\lb F,W_{i_n}\rb_{k_n}\,.
\end{align}
Here $\lb F,W\rb_n$ is a nested commutator of Eq.(\ref{commutator})
\begin{align}\label{commutator1}
&\lb F,W\rb_n=\overbrace{\big[F,[F,\cdots[F}^n,W]\cdots]],\\
&\lb F,W\rb_0=W.
\end{align}
Using (\ref{comm}) in (\ref{comm0}) and repeating the commutation of the operator exponent $e^{-\tau'_{n-2}F}$ this time with $W_{i_{n-1}}\widetilde W_{i_n}$ we get
\begin{align}\label{comm1}
&\prod\limits_{m=1}^n e^{-\tau'_{m-1}F}\,W_{i_m}\nonumber\\
&\quad=\bigg[\prod\limits_{m=1}^{n-2} e^{-\tau'_{m-1}F}\,W_{i_m}\!\bigg]\,\widetilde W_{i_{n-1}}
\,e^{-(\tau'_{n-2}+\tau'_{n-1})F},
\end{align}
where $\widetilde W_{i_{n-1}}$ expresses via $\widetilde W_{i_n}$ as
\begin{align}
&\widetilde W_{i_{n-1}}=\sum\limits_{k_{n-1}=0}^\infty\frac{(-\tau'_{n-2})^{k_{n-1}}}{k_{n-1}!}\big[ F,W_{i_{n-1}}\widetilde W_{i_n}\big]_{k_{n-1}}
\end{align}
This is the recurrent equation for the set of $\widetilde W_{i_m}\equiv\widetilde W_{i_m,i_{m+1},\ldots i_n}$, which according to the discussion above under the sum over $i_m,i_{m+1},\ldots i_n$ become {\em local differential operators}. These operators, obtained by a consecutive commutation of all operator exponents to the right, are starting from $m=n$ and terminating at $m=1$. At $m=1$ this sequence yields the expression for (\ref{comm0}) with all such exponents standing to the right of $\widetilde W_{i_1,\ldots i_n}$,
\begin{align}\label{comm1prime}
&\prod\limits_{m=1}^n e^{-\tau'_{m-1}F}\,W_{i_m}=\widetilde W_{i_1,\ldots i_n}\,e^{-\sum\limits_{m=0}^{n-1}\tau'_m F}.
\end{align}
The solution of this recurrent equation for $\widetilde W_{i_1,\ldots i_n}$ is obvious
\begin{align}
&\widetilde W_{i_1,\ldots i_n}\;\;=\!\!\!\!\!\sum\limits_{\;\;k_1,\ldots k_n=0}^\infty\!\!\!
\frac{(-\tau'_0)^{k_1}\cdots(-\tau'_{n-1})^{k_n}}{k_1!\cdots k_n!}\nonumber\\
&\quad\times\underbrace{\Big[\, F,W_{i_1}\cdots\Big[\, F,W_{i_{n-1}}}_{n-1}\Big[\, F,W_{i_n}\Big]_{k_n}\Big]_{k_{n-1}}\ldots\Big]_{k_1}.
\end{align}
Using this in (\ref{W6}) we finally get the expression for a generic order of the heat kernel expansion in terms of the operator exponent of the minimal operator $F$ with the combined overall proper time parameter $\sum_{m=0}^{n-1}\tau'_m$, standing to the right of all local differential operator coefficients
\begin{widetext}
\begin{align}\label{final}
\mathbb{K}_s^{(\beta,n)}(\tau)
=\int\limits_0^\tau d^n\tau\!\!\sum\limits_{i_0,i_1,\ldots i_n}\int\limits_{\beta\tau_n}^{\lambda_{i_0}\tau_n}\!d\tau'_0\!\!
\int\limits_{\beta\Delta\tau_1}^{\lambda_{i_1}\Delta\tau_1}\!\!\!d\tau'_1\cdots\!\!\! \int\limits_{\beta\Delta\tau_n}^{\lambda_{i_n}\Delta\tau_n}\!\!\!d\tau'_n
\!\!\!\!\!\sum\limits_{\;\;k_1,\ldots k_n=0}^\infty\!\!\!\frac{(-\tau'_0)^{k_1}\cdots(-\tau'_{n-1})^{k_n}}{k_1!\cdots k_n!}
\nonumber\\
\times \;\,K_{i_0}(\tau_n,\tau'_0)\;\mathbb{C}_{i_1\cdots i_n,k_1\cdots k_n}(\tau,\tau_1,\cdots\tau_n\,|\,\tau'_1,\cdots\tau'_n)\; e^{-\sum\limits_{m=0}^{n}\tau'_m F},
\end{align}
where we introduced the following nested $\big(\sum_{m=1}^n k_m\big)$-fold commutator
\begin{align}
\nonumber\\
&\mathbb{C}_{i_1\cdots i_n,k_1\cdots k_n}(\tau,\tau_1,\cdots\tau_n\,|\,\tau'_1,\cdots\tau'_n)\nonumber\\
&\qquad\qquad=\underbrace{\Big[\, F,W_{i_1}(\Delta\tau_1,\tau'_1)\cdots\Big[\, F,W_{i_{n-1}}(\Delta\tau_{n-1},\tau'_{n-1})}_{n-1}\Big[\, F,W_{i_n}(\Delta\tau_n,\tau'_n)\,\Big]_{k_n}\Big]_{k_{n-1}}\!\!\!\cdots\Big]_{k_1},\label{C}\\
&\Delta\tau_m\equiv\tau_{m-1}-\tau_m,\quad m=1,\cdots n,\quad \tau_0=\tau.
\end{align}
\end{widetext}

This expression is rather complicated, but it gives a systematic method of local curvature expansion or the expansion in powers of background dimensionality $\nabla^l\mathfrak{R}^m=O[1/l^{l+2m}]$. In view of background dimensionalities of the local operators $K_i(\tau,\tau')=O[1/l^0]$ and $W_i(\Delta\tau,\tau')=O[1/l]$ in the multiple commutator (\ref{C}) each perturbation order is $\mathbb{K}_s{}^{\!\!(\beta,n)}(\tau)=O[1/l^n]$. Of course, calculation with the accuracy $O[1/l^n]$ requires to consider all $\mathbb{K}_s{}^{\!\!(\beta,m)}(\tau)$ with $m=0,1,\cdots n$, each of them obtained up to terms $O[1/l^n]$ inclusive. This is achieved by substituting for minimal operator exponents the Schwinger-DeWitt expansion (\ref{DW}) and noting that $\hat a_m(F\,|\,x,x')=O[1/l^{2m}]$ along with the derivatives $\nabla_{a_1}\cdots\nabla_{a_n}\sigma(x,x')=O[1/l^{n-2}]$.

Multiple proper time integrals in (\ref{final}) have a structure similar to those of the previous section, see (\ref{sample_integral}). Because of delta function type terms in $K_i(\tau,\tau')$ and $W_i(\Delta\tau,\tau')$ some of them reduce to the contribution of the lower limit of integration over $\tau'$, and generate proper time combinations in operator exponents similar to those of the second line of Eq.(\ref{first}). For separated points they represent rather complicated multiple integrals of incomplete gamma functions generalizing (\ref{Igamma}) and (\ref{sample_integral}), but in the coincidence limit case they become elementary and yield explicitly calculable coefficients of curvature tensor structures $\nabla^l\mathfrak{R}^m$.

\section{Conclusions}
To summarize our results, we constructed a systematic method of calculating the local curvature and background fields expansion for heat kernels of nonminimal differential operators in causal theories. Causality of the theory implies that the principal symbol of its wave operator---the Hessian of the classical action---is Lorentz covariant and guarantees that the characteristic surfaces of propagating field modes coincide with the light cone in the curved spacetime metric. This method has a functorial nature, because it allows one to express the needed expansion in terms of the Schwinger-DeWitt coefficients of the auxiliary minimal operator associated with the nonminimal operator in question.

Critical starting point of the method is the knowledge of the eigenvalues of the principal symbol of the operator and the projectors on their eigenspaces. Then the method consists in several steps---construction of the leading order approximation (\ref{LObeta}) in {\em curved} spacetime, based on a special subtraction scheme (\ref{beta_subtraction}) that allows one to avoid infrared divergences associated with the infinite range of the proper time integration, the use of the noncommutative pseudodifferential calculus needed for the construction of a linear, local differential, map from the auxiliary minimal operator to the nonminimal one (\ref{final})-(\ref{C}). Finally, this is a direct use of the  Schwinger-DeWitt series to be term by term acted upon by special differential operators and multiply integrated over the proper time parameters.

The method was illustrated on the examples of the vector field operator in curved spacetime with the nondegenerate symbol and the Proca model operator with the degenerate symbol. It was shown that within the suggested method the operators with nondegenerate principle symbols have smooth heat kernels in coordinate space, while for degenerate symbols they are likely to be generalized functions with coincidence limit singularities, thus confirming that these operators go beyond the Gilkey-Seeley theory of elliptic operators.

The advantage of our approach, in distinction from numerous studies of particular nonminimal operators \cite{Gusynin1991,GusyninGorbar,GusyninGorbarRomankov, GilkeyBransonFulling,BransonGilkeyPierzchalski,GuendelmanLeonidovNechitailoOwen1994,AlexandrovVassilevich1996,AvramidiBranson2001,Avramidy2004,MossToms2014,IochumMason2017,
GrassoKuzenko2023}, is its manifest covariance in curved space fiber bundle, which is usually destroyed by the use of flat space Fourier method. Another advantage is achieved by avoiding a convolution (\ref{convolution}) of heat kernels, which is replaced by the commutator algebra (\ref{C}) of local differential operators directly leading to the needed covariant curvature expansion. As we see, the efficiency of this commutator method, demonstrated in \cite{BKW2024} for higher order minimal operators, extends to nonminimal ones. On the other hand, despite a visible simplicity of pseudodifferential method for vector field operators of Sects. \ref{Proca_section} and \ref{Vector_section}, full fledged applications in high-energy physics would certainly require a complicated perturbation theory. But the systematic nature of this technique makes it ready for a computer codification by symbolic manipulation programs.

Among the limitations of the suggested technique we can mention that it was explicitly demonstrated on the examples of second-order operators. As shown in \cite{Wach3,BKW2024} heat kernels of higher-derivative minimal operators can be systematically built in terms of the Schwinger-DeWitt expansion, so that our technique can be directly extended to the case of $N>2$ in Eq.(\ref{A2}), $\hat F(\nabla)=(-\Box)^N+\cdots$. However, the issue of a smooth coincidence limit of the heat kernel and its singularity at $\tau\to 0$ in this case remains open---$e^{-\tau F}\delta(x,x')\,|_{\,x'=x}$ has a a good coincidence limit with a finite order negative power of $\tau$, but for nonminimal operators the factor of $e^{-\tau F}\delta(x,x')$ gets numerously differentiated in Eq.(\ref{operator_structure}). This might entail infinite series of negative powers of $\tau$ which are absent in the coincidence limit of the undifferentiated kernel $e^{-\tau F}\delta(x,x')\,|_{\,x'=x}$ \cite{Wach3,BKW2024}.

Another limitation is the fact that we considered the case when all components of the matrix  valued nonminimal $\hat H(\nabla)$ have one and the same highest order in derivatives. The alternative case, involving in the principal symbol of $\hat H(\nabla)$ a number of dimensional parameters, might require the use of several different auxiliary operators $F(\nabla)$. These issues are the subject of the further research.

\section*{Acknowledgments}
Authors are grateful to Alexander Kurov as well as to an anonymous referee for valuable comments and suggestions for improving this work.
This work was supported by the Russian Science Foundation Grant No. 23-12-00051, \url{https://rscf.ru/en/project/23-12-00051/}.

\appendix

\section{Derivation of perturbation operator \label{A}}
Here we derive the expressions (\ref{Ws}) and (\ref{Wsbeta}) in the perturbation scheme of Sect.\ref{Perturbation_section} and the $\beta$-modification of this scheme in Sect.\ref{Alternative}. Also we give here the proof of Eq.(\ref{LObeta_dependence}).

The use of the heat equation in both parts of the expression (\ref{LO}) leads to the results
\begin{widetext}
\begin{align}\label{}
&\Big(\frac\partial{\partial\tau}+H\Big)\sum\limits_{k=0}^{n-1}\frac{(-\tau)^k}{k!}H^k
=\frac{(-\tau)^{n-1}}{(n-1)!}H^n=(-1)^{n-1}\int d^{n-1}\tau\,H^n,\\
&\Big(\frac\partial{\partial\tau}+H\Big)\int\limits_0^\tau d^n\tau\,
\sum_{i}\varPi_{i}(-F\lambda_{i})^n\,e^{-\tau_n\lambda_{i}F}=\int d^{n-1}\tau\,\sum_{i}\Big[\varPi_{i}(-F\lambda_{i})^n+H\varPi_{i}(-F\lambda_{i})^{n-1}\Big]\,e^{-\tau_{n-1}\lambda_{i}F}\nonumber\\
&\qquad\qquad\qquad\qquad\qquad\qquad\qquad\qquad\qquad-\int d^{n-1}\tau\,\sum_{i}H\,\varPi_{i}(-F\lambda_{i})^{n-1}.
\end{align}
Summing these expressions up we get to the equation (\ref{Wn-1}).

A similar procedure applied to the both parts of the $\beta$-modified operator (\ref{LObeta}) yields a larger number of commutator terms,
\begin{align}\label{}
&\Big(\frac\partial{\partial\tau}+H\Big)\sum\limits_{k=0}^{n-1}\frac{\tau^k(\beta F-H)^k}{k!} e^{-\beta\tau F}
=-\bigg(\frac{\tau^{n-1}(\beta F-H)^n}{(n-1)!}+\Big[\,\sum\limits_{k=0}^{n-1}\frac{\tau^k(\beta F-H)^k}{k!},\beta F\,\Big]\bigg) e^{-\beta\tau F}
\end{align}
and
\begin{align}\label{}
\Big(\frac\partial{\partial\tau}+H\Big)
\sum_{i}\int\limits^{\lambda_i\tau}_{\beta\tau} d^n\tau\,\varPi_{i}(-F)^n\,e^{-\tau_n F}
=(-1)^n\sum_{i}\int\limits^{\lambda_i\tau}_{\beta\tau} d^{n-1}\tau\,\Big[\varPi_{i}\lambda_{i}F-H\varPi_{i}\Big]F^{n-1}\,e^{-\tau_{n-1}F}\nonumber\\
+\sum_{i}\big[\,\beta\varPi_{i}F-H\varPi_{i}\big]\frac{\big[\tau F(\beta-\lambda_{i})\big]^{n-1}}{(n-1)!}e^{-\beta\tau F}.
\end{align}
Assembling them together we get (\ref{Wsbeta}).

Regarding the proof of the relation (\ref{LObeta_dependence}), direct calculation of its left hand side gives
\begin{align}\label{}
&\frac\partial{\partial\beta}\mathbb{K}^{(\beta)}_s(\tau)=\sum\limits_{k=1}^{s}\frac{\tau^k}{k!}\bigg[
-k(\beta F-H)^{k-1}F+\sum\limits_{m=0}^{k-1}(\beta F-H)^m F(\beta F-H)^{k-1-m}\bigg]\,e^{-\beta\tau F}\nonumber\\
&\qquad\quad-\frac{\tau^{s+1}}{s!}\bigg[(\beta F-H)^{s}F-\sum\limits_i\varPi_i(\beta-\lambda_i)^{s}F^{s+1}\bigg] e^{-\beta\tau F},
\end{align}
where both groups of terms in square brackets reduce to commutators or become $O[\,\mathfrak{R}\,]$ in view of the relation (\ref{H^k}). This confirms (\ref{LObeta_dependence}).
\end{widetext}

\bibliography{Wachowski2509}
\end{document}